\documentclass[final,1p]{elsarticle}

\usepackage{amssymb}

\usepackage{commath}

\usepackage[titletoc,title]{appendix}
\usepackage[utf8]{inputenc}

\DeclareFixedFont{\ttb}{T1}{txtt}{bx}{n}{12} 
\DeclareFixedFont{\ttm}{T1}{txtt}{m}{n}{12}  

\usepackage{color}
\definecolor{deepblue}{rgb}{0,0,0.5}
\definecolor{deepred}{rgb}{0.6,0,0}
\definecolor{deepgreen}{rgb}{0,0.5,0}

\usepackage{mathabx,graphicx}

\usepackage{listings}
\usepackage{quantum}
\usepackage{algorithm}
\usepackage[noend]{algpseudocode}

\PassOptionsToPackage{hyphens}{url}\usepackage{hyperref}

\newcommand\Algphase[1]{%
\vspace*{-.7\baselineskip}\Statex\hspace*{\dimexpr-\algorithmicindent-2pt\relax}\rule{\textwidth}{0.4pt}%
\Statex\hspace*{-\algorithmicindent}\textbf{#1}%
\vspace*{-.7\baselineskip}\Statex\hspace*{\dimexpr-\algorithmicindent-2pt\relax}\rule{\textwidth}{0.4pt}%
}

\newcommand\pythonstyle{\lstset{
language=Python,
basicstyle=\ttm\color{deepred},
otherkeywords={self},             
keywordstyle=\ttb\color{deepblue},
emph={MyClass,__init__},          
emphstyle=\ttb\color{deepred},    
stringstyle=\color{deepgreen},
frame=tb,                         
showstringspaces=false            %
}}

\lstnewenvironment{python}[1][]
{
\pythonstyle
\lstset{#1}
}
{}

\newcommand\pythoninline[1]{{\pythonstyle\lstinline!#1!}}
\journal{Computer Physics Communications}

\begin{document}
\sloppy
\lstset{language=Python}
\begin{frontmatter}



\title{Quintuple: a Python 5-qubit quantum computer simulator to facilitate cloud quantum computing}


\author[a,b]{Christine Corbett Moran\corref{author}}
\cortext[author] {Corresponding author.\\\textit{E-mail address:} corbett@tapir.caltech.edu}
\address[a]{NSF AAPF California Institute of Technology, TAPIR, 1207 E. California Blvd. Pasadena, CA 91125}
\address[b]{University of Chicago, 2016 SPT Winterover Scientist, Amundsen-Scott South Pole Station, Antarctica}
\begin{small}

\begin{abstract}

In May 2016 IBM released access to its 5-qubit quantum computer to the scientific community, its ``IBM Quantum Experience''\cite{IBM} since acquiring over 25,000 users from students, educators and researchers around the globe. In the short time since the ``IBM Quantum Experience'' became available, a flurry of research results on 5-qubit systems have been published derived from the platform hardware \cite{Devitt2016,Alsina2016,2016arXiv160508922R,2015arXiv151100267B,2016arXiv160501351T}. \textbf{Quintuple} is an open-source object-oriented Python module implementing the simulation of the ``IBM Quantum Experience'' hardware. \textbf{Quintuple} quantum algorithms can be programmed and run via a custom language fully compatible with the ``IBM Quantum Experience'' or in pure Python. Over 40 example programs are provided with expected results, including Grover's Algorithm and the Deutsch-Jozsa algorithm. \textbf{Quintuple} contributes to the study of 5-qubit systems and the development and debugging of quantum algorithms for deployment on the ``IBM Quantum Experience'' hardware.





\end{abstract}
\end{small}


\end{frontmatter}



{\bf PROGRAM SUMMARY}

\begin{small}
\noindent
{\em Manuscript Title:}  Quintuple: a Python 5-qubit quantum computer simulator to facilitating cloud quantum computing                        \\
{\em Authors: } Christine Corbett Moran                                                \\
{\em Program Title: } Quintuple                                         \\
{\em Licensing provisions: none}                                   \\
{\em Programming language: } Python                                  \\
{\em Computer: } Any which supports Python 2.7+                                               \\
{\em Operating system: } Cross-platform, any which supports Python 2.7+, e.g. Linux, OS X, Microsoft Windows                                       \\
{\em RAM:} 200 Mb                                              \\
{\em Number of processors used:} 1                              \\
{\em Supplementary material:}   Manual available at \url{https://github.com/corbett/QuantumComputing} 
{\em Classification: 4.15}                                         \\
{\em External routines/libraries: } Numpy (\url{http://numpy.scipy.org})                          \\
{\em Nature of problem:} IBM has released access to a 5-qubit quantum computer via its ``IBM Quantum Experience''[1]. Classical simulations on 5-qubit systems can provide insight into the function and performance of quantum algorithms and aid students and educators in their study. Developing and debugging algorithms for deployment on the IBM Quantum Experience can be assisted by a custom simulation infrastructure compatible with its hardware.\\
{\em Solution method:}
\textbf{Quintuple} provides an open-source object-oriented 5-qubit quantum computer class in the widely used Python language, with full support for all operations available on the IBM Quantum Experience hardware. This quantum computer class can be used interactively or scripted, in native python or using a simplified syntax directly compatible with that used on the IBM Quantum Experience.  \\
{\em Restrictions:}
  \textbf{Quintuple} is implemented for simulations of up to 5-qubits and is designed to support the gates and syntax available on the IBM Quantum Experience hardware. \textbf{Quintuple} is designed flexibly such that it can easily be extended to support further qubits, gates, syntax, and algorithmic abstractions as the IBM Quantum Experience hardware expands in functionality to keep parity. \\
{\em Running time:}
Typical running time the execution of a non-trivial quantum algorithm and comparing its output to an expected output is on the order of a thousandth to a hundredth of a second.
   \\

\end{small}

\section{Introduction}
Quantum computers can perform certain tasks more efficiently than classical computers \cite{Shor:1997:PAP:264393.264406,Bennett1997}. Furthermore, the results and limitations of realistic quantum computers gives us insight into the fundamentals of quantum mechanics. Quantum computation has thus attracted great interest from the research community. Recently IBM has released access to its 5-qubit quantum computer to the scientific community under the moniker ``IBM Quantum Experience''\cite{IBM}. The IBM Quantum Experience provides access to a 5-qubit quantum computer with a limited set of gates described by IBM as ``the world’s first quantum computing platform delivered via the IBM Cloud''. A body of research focuses on properties of 5-qubit systems \cite{Touchette2010,Das2004}, and much of it has recently been released or updated to rely upon results running on IBM's Quantum Experience \cite{Devitt2016,Alsina2016,2016arXiv160508922R,2015arXiv151100267B,2016arXiv160501351T}. Understanding the capabilities of, and developing and debugging algorithms for deployment on this infrastructure, calls for analyzing and simulating 5-qubit systems in detail.

A variety of existing software toolkits are useful in quantum computation study and research, ranging from the general QuTIP \cite{Johansson2012,Johansson2013} available in Python, to more specialized toolkits available in a variety of scientific computing languages: QUBIT4MATLAB (Matlab)\cite{Toth2008}, QCMPI (Fortran 90)\cite{Tabakin2009} provide rapid evaluation of quantum algorithms, including noise analysis, for a large number of qubits by exploiting parallel computing. The FEYNMAN (Maple) \cite{Radtke2005,Radtke2006,Radtke2007,Radtke2008} program offers interactive simulations on $n$-qubit quantum registers without restrictions other than available memory and time resources of computation. The QDENSITY (\textit{Mathematica})\cite{Julia-Diaz2009} program provides commands to create and analyze quantum circuits. The {\tt libquantum} package (C) provides the ability to simulate a variety of processes based on its implementation of a quantum register\cite{libquantum}, Qinf (Maxima) allows the manipulation of instances of objects that appear in quantum information theory and quantum entanglement\cite{qinf}. A detailed comparison between other quantum simulators is beyond the scope of this work. Those that are available use various computer languages, the majority in C/C++, and have different focuses, ranging from particular algorithms, generalisability, or scalability. A more comprehensive list of available tools for work in quantum computation is given on the Quantiki wiki\cite{availablesims}.

In this paper, I describe \textbf{Quintuple}, an open-source Python module allowing both simulation of all operations available via IBM's Quantum Experience hardware and programming for a 5-qubit quantum computer at a high level of abstraction\cite{quintuplegit}. \textbf{Quintuple} allows the researcher, educator or student to quickly and repeatedly execute code in a simplified language compatible with execution on the IBM Quantum Experience hardware and/or in pure Python compatible with other Python code and libraries. By dialing in the focus of \textbf{Quintuple} on the uniquely available IBM Quantum Experience hardware, it can be deployed on the platform without additional configuration. By keeping the implementation to just those elements necessary to perform an ideal simulation of IBM's 5-qubit quantum computer, and not relying on a much larger, fuller featured toolkit, as well as by providing an open-source object-oriented implementation in a widely used high level language, Python, it is hoped this module will be useful to more novice programmers and/or those less experienced in the intricacies of quantum computation. The core of \textbf{Quintuple} is only 675 lines of Python code, and \textbf{Quintuple} additionally provides over 40 example programs with expected results, including examples of Grover's Algorithm and the Deutsch-Jozsa algorithm, for execution within \textbf{Quintuple} or on the IBM Quantum Experience.

 In \textbf{Section \ref{quantuminf}} a brief overview is given of the terminology and mathematics necessary to follow the operation of \textbf{Quintuple}. In \textbf{Section \ref{quintuple}}, the \textbf{Quintuple} code and the APIs to design and test 5-qubit quantum algorithms in simulation and/or on IBM's hardware are introduced. In \textbf{Section \ref{usage}}, through the lens of an algorithm which swaps the state of two qubits, various modes of usage of the APIs presented in \textbf{Section \ref{quintuple}} are presented. \textbf{Section \ref{conclusion}} provides a summary and outlook for potential future work.

\section{Overview of Quantum Information}
\label{quantuminf}
Here I give a brief primer on quantum information and computation necessary to describe \textbf{Quintuple}'s implementation and assisting in understanding the IBM Quantum Experience. Knowledge of complex conjugation, basic linear algebra fluency; matrix operations including multiplication, tranpose, trace, and tensor products is assumed, among other mathematical concepts. Sexplicit exposition of this formalism and any explanation of the \emph{whys} of quantum mechanics is beyond the scope of this limited overview of quantum information. For an detailed overview of the math and quantum mechanics of quantum information, as well as a lucid exposition of the fundamentals of quantum information in detail, an excellent resource is the canonical textbook of Nielsen and Chuang\cite{nielsen2010quantum}. For further overview of the simulation of $n$-qubit systems the overview by Radtke is an excellent supplement to this more limited exposition\cite{Radtke2005}.

A qubit is the quantum generalization of a classical bit. Unlike a classical bit, it can take any value corresponding to a linear superposition of its constituents: formally two orthonormal eigenstates. Our default choice of basis throughout this manuscript is $$ \{ \ket{0}=\begin{pmatrix}1 \\ 0\end{pmatrix},\ket{1}=\begin{pmatrix}0 \\ 1\end{pmatrix} \}$$. This multi-purpose notation ($\bra{}$ or $\ket{}$), used throughout this manuscript to represent a quantum state, is called bra-ket or Dirac notation and is standard in quantum mechanics. Without getting into a detailed discussion of the mathematics, one can, simplistically, think of the symbol lying between the $\ket{}$ notation as being a label for the state. Whether the notation is $\ket{symbol}$ vs. $\bra{symbol}$ indicates whether it is represented as a column or a row vector respectively, where $\bra{symbol}$ is the conjugate transpose of $\ket{symbol}$ and vice versa.  

Thus a generic one-qubit state $\ket{\psi}$ is
\begin{equation}
\ket{\psi}=a\ket{0}+b\ket{1}.
\end{equation}
The coefficients $a,b$ are complex numbers and these complex coefficients provide the representation of $\psi$ in the ${\ket{0},\ket{1}}$ basis. The probability of finding $\ket{\psi}$ in state $\ket{0}$ is $\abs{a}^2=aa^*$ where $a^*$ is the complex conjugate of $a$, similarly the probability of finding $\ket{\psi}$ in state $\ket{1}$ is $\abs{b}^2=bb^*$. These two probabilities normalize to one: $\abs{a}^2 + \abs{b}^2=1$. A single qubit state $\ket{\psi}$ can be physically realized by a variety of mechanisms which correspond to a quantum-mechanical two-state systems, for example a two spin system, or a two level system, among many others. The Bloch sphere is a useful way to visualize the state of a single qubit on a unit sphere. Formally, in the Bloch sphere representation the qubit state is written as
\begin{equation}
\ket{\psi} = \cos{\frac{\theta}{2}}\ket{0}+e^{i\phi} \sin{\frac{\theta}{2}}\ket{1},
\end{equation}
where $\theta$ and $\phi$ are the polar coordinates to describe a vector on the unit sphere. 

To make use of the power of quantum computation we will in general want more than one qubit. In a classical $n$-bit register we can initialize each bit to 0 or 1. For example to represent the base 10 number 19 in a classical 5-bit register we can set its elements to $10011$. For $n$ qubits, to create an analogous state, a so-called quantum register we prepare the state $\ket{10011}=\ket{1} \otimes \ket{0} \otimes \ket{0} \otimes \ket{1} \otimes  \ket{1}$. Here $\otimes$ corresponds to the tensor product (also known as the direct or Kronecker product). Generically an $n$-bit quantum register can hold any superposition of $n$-qubit states.

For an $n$ qubit state there are $2^n$ possible values of which the $n$-qubit state can, in general, be a superposition of. For example for a 2-qubit state we have $2^2=4$ possible states, $\{\ket{00},\ket{01},\ket{10},\ket{11}\}$. For a 3-qubit state we have $2^3=4$ possible states or 
\begin{equation}
 \{\ket{000},\ket{001},\ket{010},\ket{011},\ket{100},\ket{101},\ket{110},\ket{111}\}.
\end{equation}

Numbering the states from $0$ to $2^n-1$, the canonical ordering used throughout this manuscript is:
\begin{equation}
\label{canonicalordering}
\sum_{m=0}^1 \ldots \sum_{j=0}^1   \sum_{i=0}^1 \ket{i j ... m },
\end{equation}
 where the number of summations corresponds to the total number of qubits. Thus if we incorporate the amplitudes, the complex coefficients of these states, we can compute the probability of finding 
\begin{equation}
\ket{\psi}=\sum_{m=0}^1 \ldots \sum_{j=0}^1   \sum_{i=0}^1 c_{ij\ldots m}\ket{i j ... m },
\end{equation}
in state $\ket{i j ... m }$ as the squared absolute value of $c_{ij\ldots m}$, $|c_{ij\ldots m}|^2=c_{ij\ldots m} c_{ij\ldots m}^*$.

If we can represent an $n$-qubit state as the tensor product of the states of individual qubits 
\begin{equation}
\ket{q_0 q_1 \ldots q_n}=\ket{q_0} \otimes \ket{q_1} \otimes \ldots \otimes \ket{q_n},
\end{equation}
the state is called separable. However, due to the nature of superposition, it may be that a multi-qubit state is non-separable and individual qubits states are not well defined independent of other qubits.  This non-local correlation phenomenon known as entanglement is a necessary resource to achieve the exponential speed up of quantum compared to classical computation \cite{jozsa2003role}. As such, the concept of quantum registers, necessary to store multi-qubit non-separable states, will play a primary role in quantum computation simulation.

We have outlined the analogy to the classical n-bit register, the n-qubit quantum register for keeping track of quantum data. Here we will do the same with a classical gate and a quantum gate, which evolve classical and quantum states respectively.  In classical computation,  a classical gate operates on a classical register to evolve its state.  In quantum computation, a quantum gate operates on a quantum register to evolve its state. Quantum states can be represented by matrices; the mathematics of the evolution of quantum states can unsurprisingly be represented by matrices as well. To represent quantum gates, these matrices must conform with the postulates of quantum mechanics as they multiply a state to produce an evolved state. Specifically, we know that the evolution of states must conserve probability (preserve norms); we cannot produce a state which is a superposition of states with probability greater than one. 

Matrices which ensure the conservation of probability when they multiply states are called \textit{unitary}.  Formally, this corresponds to any matrix $U$ which satisfies the property that its conjugate transpose $U^\dagger$ is also its inverse, that is $U^\dagger U=UU^\dagger=I$, where $I$ is the identity matrix. In quantum computation, a quantum gate corresponds to a unitary matrix, and any unitary matrix corresponds to a valid quantum gate. Since unitary matrices are always invertible, quantum gates and thus computation is reversible; any operation we can do we can undo\cite{bennett1973logical}.  As a qubit state can be realized physically by a variety of quantum mechanical systems, so can quantum gates be physically realized by a variety of quantum mechanical mechanisms, which must necessarily depend on the system's representation of the qubit. For example, in a system where qubits are represented by ions in a quantum trap, a laser tuned to a particular frequency can induce a unitary transformation effectively acting as a quantum gate. 

Gates acting on a single qubit can be applied to a quantum register of an arbitrary qubit number.
For example, for a gate X if the desired qubit to act on is the 3rd qubit in a 4-qubit quantum register. X is a gate which flips the qubit it acts on from $\ket{0}$ to $\ket{1}$ or from $\ket{1}$ to $\ket{0}$. The appropriate gate is formed via
$X_{3 \text{of} 4}=I \otimes I \otimes X \otimes I $ where $I$ is the $2 \times 2$ identity matrix. In general, to create a gate $G_{m \text{of} n}$ to operate on the $m$th qubit of a register of $n$ qubits from a gate $G$ that operates on a single qubit, one may use 
\begin{equation}
G_{i \text{ of } n} = \bigotimes_{i=1}^{n} 
\begin{cases} 
I & \text{ if } i \neq m \\
G & \text{ if } i=m.
\end{cases}
\end{equation}
Here, $\bigotimes_{i=1}^{n}$ is the analog of $\sum_{i=0}^{n}$ corresponding to the tensor product, instead of the summation operation. We can see that the application of a gate on a single qubit in this fashion doesn't generate entanglement as it never results in the expansion of the size of the quantum register it is acting on.

Specific sets of classical gates, for example the the NOT and AND gates can be used to construct all other classical logic gates and thus forms a set of universal classical gates. Other such sets exist; in fact the NAND, \textit{negative and} gate alone is a universal classical gate\cite{nielsen2010quantum}. In quantum computation, to obtain a universal gate set we will need a multi-qubit gate which applies on 2-qubits of an $n$-qubit register. The CNOT gate is one such gate. CNOT is the \textit{2-qubit controlled not gate}. Its first input is known as the control qubit, the second as the target qubit and the state of the target qubit is flipped on output if and only if the control qubit is $\ket{1}$. The application of CNOT can under many scenarios generate entanglement. CNOT combined with single qubit gates can approximate arbitrarily well any (unitary) operation on a quantum computer\cite{Barenco1995}. Quantum gates can be combined to form quantum circuits, the analog to classical circuits composed of logic gates connected by wires. The full set of gates that both the IBM Quantum Experience and \textbf{Quintuple} support, form a (non-minimal) universal quantum gate set, such that we can combine the gates in a quantum circuit to create any multi-qubit logic gate we desire. 

We'll need to understand how measurement functions in quantum mechanics to understand the constraints of extracting information from a quantum register. Measurement in quantum mechanics is something which engages a lot of discussion, but its properties are straightforward to state in mathematics if not in philosophy.  It is possible to perform a measurement of a single qubit with respect to any basis $\{\ket{a},\ket{b}\}$ (not just the default $\{\ket{0},\ket{1}\}$ basis) so long as this basis is orthonormal, that is that the total probability is one. It likewise is possible to measure a multi-qubit system with respect to any orthonormal basis. Earlier, we stated that the probability of finding:
\begin{equation}
 \ket{\psi}=\sum_{m=0}^1 \ldots \sum_{j=0}^1   \sum_{i=0}^1 c_{ij\ldots m}\ket{i j ... m },
\end{equation}
in state $\ket{i j ... m }$ is the squared absolute value of $c_{ij\ldots m}$, $\abs{c_{ij\ldots m}}^2=c_{ij\ldots m} c_{ij\ldots m}^*$. Here when we perform a measurement we actually do find the system in one of these states $\ket{i j ... m}$ with the appropriate probability  $|c_{ij\ldots m}|^2$. After the measurement is performed, the state is collapsed and all further measurements return the same result, state $\ket{i j \ldots m}$ with probability 1.

\section{Quantum information tools represented in Quintuple}
\label{quintuple}

Only those states and gates which are useful to interfacing with IBM's 5-qubit quantum computer are supported by \textbf{Quintuple}. The only external Python module \textbf{Quintuple} relies upon is the \textbf{numpy} module. The core of \textbf{Quintuple} is just 675 lines long. 
\subsection{States}
States are available as static member variables of the \pythoninline{class State}. The following qubit states are are available

Standard (z) basis (\pythoninline{State.zero_state,State.one_state}):
\begin{equation}
\ket{0}=\begin{pmatrix} 1 \\ 0\end{pmatrix}, \ket{1}=\begin{pmatrix}0 \\ 1 \end{pmatrix}.
\end{equation}

Diagonal (x) basis (\pythoninline{State.plus_state,State.minus_state}):
\begin{equation}
 \ket{+}= {\frac{1}{\sqrt{2}}} \begin{pmatrix} 1 \\ 1 \end{pmatrix}, \ket{-}=\frac{1}{\sqrt{2}}\begin{pmatrix}1 \\ -1 \end{pmatrix}.
\end{equation}

Circular (y) basis (\pythoninline{State.plusi_state,State.minusi_state}):
\begin{equation}
 \ket{\circlearrowright}= {\frac{1}{\sqrt{2}}} \begin{pmatrix} 1 \\ i \end{pmatrix}, \ket{\circlearrowleft}=\frac{1}{\sqrt{2}}\begin{pmatrix}1 \\ -i \end{pmatrix}.
\end{equation}

The \pythoninline{class State} has a variety of helper methods, including those to transform to the $x$ or $y$ basis, to see if a multi-qubit state is simply separable into individual qubits in the set $\{\ket{0},\ket{1},\ket{+},\ket{-},\ket{\circlearrowright},\ket{\circlearrowleft} \}$, and to extract the $n$th qubit from a separable multi-qubit state. This class implements the measurement method, following the limitations of nature, and supports retrieving a state's representation on the Bloch sphere, not possible in nature but feasible in simulation. The class also has a method to create a state from binary string (e.g. ``01011'' corresponding to $\ket{01011}$) and return a string from a separable state. For example, we can compute the representation of the state $\ket{10011}$ in the \textbf{Quintuple} module numerically with
\begin{python}
np.kron(State.one_state,np.kron(
  State.zero_state,np.kron(
  State.zero_state,np.kron(
  State.one_state,State.one_state))))
\end{python} or more concicely by \pythoninline{State.state_from_string("10011")}.

\subsection{Gates}
A variety of single qubit gates are supported, as is the 2-qubit gate CNOT. Later, the \pythoninline{class QuantumComputer} will use these gates as building blocks to define gates which operate on quantum registers of up to 5-qubits appropriately. In the following gate definitions the Python syntax is given in parenthesis.

$H$ gate; Hadamard gate (\pythoninline{Gate.H}):
\begin{equation}
H={\frac{1}{\sqrt{2}}}\begin{pmatrix} 1 & 1 \\ 1 & -1 \end{pmatrix}.
\end{equation}

$X,Y,Z$ gates; Pauli gates (\pythoninline{Gate.X,Gate.Y,Gate.Z}):
\begin{equation}
X=\begin{pmatrix} 0 & 1 \\ 1 & 0 \end{pmatrix},
\end{equation}
\begin{equation}
Y=\begin{pmatrix} 0 & -i \\ i & 0 \end{pmatrix}.
\end{equation}
\begin{equation}
Z=\begin{pmatrix} 1 & 0 \\ 0 & -1 \end{pmatrix}.
\end{equation}

$I$ gate; Identity gate (\pythoninline{Gate.eye}):
\begin{equation}
I=\begin{pmatrix} 1 & 0 \\ 0 & 1 \end{pmatrix}.
\end{equation}

$S$ gate; Phase gate (\pythoninline{Gate.S}):
\begin{equation}
S=\begin{pmatrix} 1 & 0 \\ 0 & i \end{pmatrix}.
\end{equation}

$S^\dagger$ gate (\pythoninline{Gate.Sdagger}):
\begin{equation}
S^\dagger=\begin{pmatrix} 1 & 0 \\ 0 & -i \end{pmatrix}.
\end{equation}

$T$ gate; $\pi/8$ gate (\pythoninline{Gate.T}):
\begin{equation}
T=\begin{pmatrix} 1 & 0 \\ 0 & e^{\frac{i \pi}{4}} \end{pmatrix}.
\end{equation}

$T^\dagger$ gate (\pythoninline{Gate.Tdagger}):
\begin{equation}
T^\dagger=\begin{pmatrix} 1 & 0 \\ 0 & e^{-\frac{i \pi}{4}} \end{pmatrix}.
\end{equation}

$CNOT$ gate (\pythoninline{Gate.CNOT2_01}):
\begin{equation}
CNOT=\begin{pmatrix} 1 & 0 & 0 & 0 \\ 0 & 1 & 0 & 0 \\ 0 & 0 & 0 & 1 \\ 0 & 0 & 1 & 0 \end{pmatrix}.
\end{equation}
It can easily be checked that these gates produce the desired behavior. All other combinations of target and control qubits are available within \pythoninline{class Gate} acting on quantum registers of up to 5 qubits. Here, the number appearing after the CNOT indicates the number of qubits in the register the gate is to operate on, the first subscript indicates the control qubit index in the entangled qubit register, and the second subscript indicates the target qubit index, both 0 based. For example \pythoninline{CNOT4_03} is to operate on a 4-qubit register with the 0th qubit corresponding to the control qubit and the 3rd qubit corresponding to the target qubit. The \pythoninline{class QuantumComputer} helpfully supports specifying only the target and control qubits when applying the CNOT gate and automatically deploys the correct gate to achieve this based on the internal configuration of its quantum registers.

\subsection{Probabilities}
Several convenience methods are provided to help compute probabilities and expectation values. For a qubit residing in a quantum register representing an arbitrary number of entangled qubits, the method \pythoninline{Probability.get_probabilities(qubit)} returns an array of probabilities representing the quantum register in the canonical ordering defined in \textbf{Equation \ref{canonicalordering}}. The method \pythoninline{Probability.pretty_print_probabilities} prints each state and its associated probability for easy examination. For a state representing a single qubit, there are several additional methods available within \pythoninline{class Probability} to help calculate the expectation of the state in the standard (z), circular (x) or diagonal bases, respectively. 

\subsection{QuantumRegister}
\label{quantum_register}

To represent a possibly non-separable group of distinguishable qubits, one can treat them together in terms of a single quantum register to keep track of their ordering and their entangled state. \textbf{Quintuple} uses the \pythoninline{class QuantumRegister} for this purpose, and the register is managed by the \pythoninline{class QuantumComputer} so that it isn't necessary to follow how the qubits within \pythoninline{class QuantumComputer} are internally arranged for the user to be able to perform operations and measurements. The \pythoninline{QuantumRegister} object can be queried as to the number of qubits it represents, which particular qubits it represents, its state, and whether it is equal to another \pythoninline{QuantumRegister} object. 

The \pythoninline{class QuantumRegister} has an additional method not provided in nature. Specifically a qubit is a superposition of states and when measured its state collapses to just one of these states with a probability given by the probability amplitude squared. All further measurements return the same state as the qubit is no longer in a superposition of states. The \pythoninline{QuantumComputer} supports measurement in the fashion of nature, but it also for convenience of further analysis, saves the value of the full state before collapse in the \pythoninline{QuantumRegister} object, which can be retrieved with the method \pythoninline{get_noop()}.

\subsection{QuantumRegisterCollection}
\label{quantum_register_collection}
The \pythoninline{class QuantumRegisterCollection} is an abstraction that assists the \pythoninline{class QuantumComputer} in managing its \pythoninline{QuantumRegister}s. This class returns the register in which a particular qubit resides, manages the merging of two \pythoninline{QuantumRegister}s under the hood via its \pythoninline{entangle_qubits} method, and allows easy querying as to the order of the qubits it is representing. This is useful to the \pythoninline{class QuantumComputer} as it supports the user querying about the state of the qubits in any increasing order the user desires. The abstraction of the \pythoninline{QuantumRegisterCollection} allows the \pythoninline{QuantumComputer} to keep the qubits separately, in separate registers for as long as possible, only merging into a single register when necessary. This means that the matrix operations associated with gate action are kept smaller and that states are kept separated for clarity for as long as is possible.

\subsection{QuantumComputer}
The \pythoninline{class QuantumComputer} manages five qubits in an arbitrary grouping of quantum registers and allows the user to apply quantum gates and measure and extract state information without having to consider how the qubits are internally represented. At creation or upon reset, the \pythoninline{class QuantumComputer} prepares five qubits named ``q0'',``q1'',``q2'',``q3'',``q4'',and ``q5'' each having state $\ket{0}$. Its two primary methods are \pythoninline{apply_gate} and \pythoninline{apply_two_qubit_gate_CNOT}, which allow the user to apply the respective one and two qubit quantum gates which \textbf{Quintuple} supports. Additionally, the \pythoninline{execute} method allows the user to execute code snippets in a simplified syntax designed to be fully compatible with execution on the IBM Quantum Experience hardware, which compiles to use the appropriate pure Python methods. After the evolution code has been executed, the internal state can be easily queried and compared to expected results.

\subsubsection{Applying Gates to Individual Qubits}
The method \pythoninline{apply_gate} takes as arguments a gate of the \pythoninline{class Gate} and the name of the qubit to act on. Under the hood, this method acts on this qubit by simply applying the gate if the qubit is the only element of its quantum register, or if the qubit is a member of a quantum register with more than one element, by creating and applying the corresponding gate to act on that qubit within the register. 

\subsubsection{Applying Controlled Gates to Two Qubits}
The \pythoninline{apply_two_qubit_gate_CNOT} method has a similar syntax, taking as an argument the name of the control and the name of the target qubit. No gate name is needed as this method handles CNOT only. The quantum register(s) containing these two qubits, potentially the same register, are found within the \pythoninline{QuantumComputer}'s \pythoninline{QuantumRegisterCollection}. If the two quantum registers the method is acting on (containing the control and the target qubit respectively) are distinct and each contain one qubit only, then a combined state corresponding to the tensor product of these two states is created and the default \pythoninline{Gate.CNOT2_01} gate is applied. If after application, the combined state is fully separable into two individual qubits in the $z$ basis ($\{\ket{0}, \ket{1}\}$), the target qubit is alone changed and the two are not entangled. If, however, after the application the combined state is not fully separable in this fashion, they are merged into a single quantum register. 

If one or both of the quantum registers given contain more than one qubit, then their states are likewise combined via a tensor product as necessary (if they don't already reside in the same quantum register). Then the appropriate CNOT matrix formulation for the combined state is applied to the combined state. The state of the relevant quantum register--the new register if one was created, otherwise the existing register which held both qubits--is set to the output of this calculation. Although \textbf{Quintuple} currently supports only the \pythoninline{CNOT} controlled gate out of the box; additional controlled gates could be easily supported. Indeed, given that the gate set \textbf{Quintuple} supports is universal further controlled gates can be built out of supported components without modification.

\subsubsection{Measurement}
The \pythoninline{measure} method of the \pythoninline{class QuantumComputer} does a probabilistic measurement of the quantum register in which the desired qubit resides. The measurement is performed in the default ($z$) basis and collapses the state. Since we are in a simulation, we can perform the same computation repeatedly and verify that the measurement operation statistically converges to the distribution given by the probability amplitude of the state in superposition resulting from the computation. Since we are in a simulation we can also have direct access to these amplitudes. For convenience, before a measurement is performed the state in superposition is stored and is accessible later via the method \pythoninline{get_noop()} of the \pythoninline{class QuantumRegister}. Nature doesn't give us this information, but the \textbf{Quintuple} module can. This is useful for testing or later analysis. The \pythoninline{bloch} method of \pythoninline{class QuantumComputer} implements the capability of visualizing a single qubit on the bloch sphere. If the value of \pythoninline{get_noop()} is set, the state has been accessed and is collapsed. 

\subsubsection{Checking Output}
Internally, \pythoninline{class QuantumComputer} may be representing qubits in any combination of \pythoninline{QuantumRegister}s and within each \pythoninline{QuantumRegister} in any order. To compare to expected outputs, we need to be able to compare the probability amplitudes or qubit states for a collection of qubits in a specified order or to compare the Bloch coordinates for a given qubit to an expected result. Thus, methods are provided so that the user can specify a qubit or group of qubits in a comma separated string, along with the expected result in their specified order and use an equality to test whether the result matches. The algorithm used to output the entangled state in the desired order is given in \textbf{Appendix \ref{reordering}}. At this time the requested order must be in increasing qubit index order due to the detailed implementation of the reordering algorithm.

The \pythoninline{probabilites_equal} and \pythoninline{qubit_states_equal} methods function similarly, the former comparing probabilities and the latter amplitudes. If one of the quantum registers contains the requested qubits in order directly, this is simply computed and returned. Otherwise, an algorithm is run to output an entangled state representing the ordered tensor product of the requested qubits, and the probability or amplitude vector representing this entangled state is compared to that specified by the user. The \pythoninline{bloch_coords_equal} simply compares the Bloch representation of the desired qubit to that specified, if it happens to be in its own quantum register. If the desired qubit is in a quantum register with other qubits, it attempts to separate it from the quantum register in which it resides. The ``easy'' separation algorithm is simplistic, and only succeeds if the state is a permutation of the tensor product single-qubit states which are in the set $\{\ket{0},\ket{1},\ket{+},\ket{-},\ket{\circlearrowright},\ket{\circlearrowleft}\}$. Thus, just because the separation algorithm returns failure does not imply the state is fundamentally inseparable. If the desired qubit is not easily separable from others in its quantum register, the comparison method raises an exception. If it is, then the method finds the desired qubit, now in a single qubit state, and compares the result to that desired.

\subsubsection{Execution of Programs in IBM's syntax}
Programs for execution on \textbf{Quintuple}'s \pythoninline{class QuantumComputer} can be written in a concise format, compatible with direct execution on the IBM Quantum Experience hardware. It is the language which is printed out to accompany the graphical setup of states, gates, and measurement operations in the IBM Quantum Experience interface. The interface also allows the user to simply copy and paste programs in this language, rather than forcing them through the graphical intermediary. 

Currently, the following syntax for use in  \pythoninline{class QuantumComputer}'s \pythoninline{execute} method  encompasses all that a user is able to do on the 5-qubit IBM Quantum Experience hardware:

\begin{tabular} {| l | r | }
\hline
available qubit list & $q[0],q[1],q[2],q[3],q[4]$ \\
1-qubit gate list & h,t,tdg,s,sdg,x,y,z,id \\
1-qubit gate action & ``gate q[i];'' \\
2-qubit CNOT gate list & cnot \\
2-qubit CNOT gate action & ``cnot q[control], q[target];'' \\
measurement operation list & measure, bloch \\
measurement operation action & ``operation q[i];'' \\
\hline
\end{tabular}

\noindent Here $\{$h,t,tdg,s,sdg,x,y,z,id$\}$ correspond to the Python
\begin{python}
Gate.H, Gate.T, Gate.Tdagger, Gate.S, Gate.Sdagger, 
  Gate.X, Gate.Y, Gate.Z, Gate.eye
\end{python} 

A program in this syntax it can be executed easily.  Program code can be put in a Python string or equivalently read in from a file into a string. The code can be executed with the \pythoninline{execute} method, and afterwards the state of the quantum computer can be probed as desired in pure Python. The following \textbf{Section \ref{usage}} contains an explicit example of code in this syntax and its usage. For convenience, although the \pythoninline{execute} method takes in a string representing the program code, for testing and keeping track of program output the \pythoninline{class Program} is provided. This has the code in its \pythoninline{code} variable but additionally can store an expected \pythoninline{result_probability} or \pythoninline{bloch_vals}. For perusal, use, elaboration and testing, over 40 example programs are collected in the \pythoninline{class Programs}. 

\section{Quintuple Code, exploration of modes of usage}
\label{usage}

In this section a variety of modes of usage of the \textbf{Quintuple} module are provided. For consistency and comparison, each example mode of usage executes the same algorithm, corresponding to swapping the states of two qubits. As a more detailed overview of the action of this algorithm, here the quantum computer begins with $\ket{q_1}=\ket{0}$, $\ket{q_2}=\ket{0}$. Then the code applies the $X$ gate to $\ket{q2}$ which inverts it to be $\ket{q_2}=\ket{1}$. Thus at the initial stage $\ket{q_1 q_2}=\ket{01}$. The algorithm then applies a series of CNOT and H gates such that we end up with $\ket{q_1 q_2}=\ket{10}$, a swapping of the states of the qubits at the initial stage.

\subsection{Syntax compatible with IBM Quantum Experience hardware}
To prepare for execution, we set the \pythoninline{swap_code} variable to the string containing the program code: 

\begin{python}
x q[2];
cx q[1], q[2];
h q[1];
h q[2];
cx q[1], q[2];
h q[1];
h q[2];
cx q[1], q[2];
measure q[1];
measure q[2];
\end{python}
\noindent We can then execute and examine the results with:
\begin{python}
qc=QuantumComputer()
qc.execute(swap_code)
Probability.pretty_print_probabilities(
     qc.qubits.get_quantum_register_containing(
     "q1").get_state())
\end{python}
\noindent which will print, as expected:
\begin{python}
|psi>=|10>
Pr(|10>)=1.000000; 
\end{python}

\subsection{Swap program in pure Python}
This same algorithm can be executed in pure python using the machinery of \pythoninline{class QuantumComputer}
\begin{python}
qc=QuantumComputer()
qc.apply_gate(Gate.X,"q2")
qc.apply_two_qubit_gate_CNOT("q1","q2")
qc.apply_gate(Gate.H,"q1")
qc.apply_gate(Gate.H,"q2")
qc.apply_two_qubit_gate_CNOT("q1","q2")
qc.apply_gate(Gate.H,"q1")
qc.apply_gate(Gate.H,"q2")
qc.apply_two_qubit_gate_CNOT("q1","q2")
qc.measure("q1")
qc.measure("q2")
\end{python}

\subsection{Swap in Pure python, without the QuantumComputer machinery}
Equivalently this algorithm can be run using the machinery of the \textbf{Quintuple} module states and gates without relying on the abstraction of its \pythoninline{class QuantumComputer} in the following manner:
\begin{python}
q1=State.zero_state
q2=State.zero_state
q2=Gate.X*q2
new_state=Gate.CNOT2_01*np.kron(q1,q2)
H2_0=np.kron(Gate.H,Gate.eye)
H2_1=np.kron(Gate.eye,Gate.H)
new_state=H2_0*new_state
new_state=H2_1*new_state
new_state=Gate.CNOT2_01*new_state
new_state=H2_0*new_state
new_state=H2_1*new_state
new_state=Gate.CNOT2_01*new_state
\end{python} 

This manner of working with the module provides the most complete mathematical understanding of the operations that \pythoninline{class QuantumComputer} is abstracting. Any individual state or gate can be printed, and it is clear how entanglement is represented as this is not done under the hood. This mode of execution also provides the most clear understanding of the convenience that \textbf{Quintuple}'s \pythoninline{class QuantumComputer} affords. Explicit execution in this manner requires a more complicated syntax, manual management of quantum registers, and no convience methods are available. 

\section{Summary and Outlook}
\label{conclusion}
\textbf{Quintuple} has been developed to aid the study and research of 5-qubit systems. \textbf{Quintuple} facilitates the development and debugging of quantum algorithms for deployment on IBM's Quantum Experience by providing an out-of-the-box self-contained ideal simulator of IBM's 5-qubit hardware and software infrastructure. Using the widely available and open source computer language Python and its numerical module \textbf{numpy}, \textbf{Quintuple} provides full support for all operations available on the IBM Quantum Experience hardware. This quantum computer class can be used interactively or scripted, in native Python or using a simplified syntax directly compatible with that used on the IBM Quantum Experience infrastructure. \textbf{Quintuple} has been designed to be flexible enough to be simply extended to support further qubits, gates, syntax, and algorithmic abstractions as the IBM Quantum Experience infrastructure itself expands in functionality. 

Several extensions of \textbf{Quintuple} are planned. First, as the IBM Quantum Experience evolves, whether to support additional gates, qubit number, or abstraction, \textbf{Quintuple} will necessarily need to be updated to keep parity. Some of these updates can and will be done in anticipation, so long as the simplicity of \textbf{Quintuple} is maintained and is backwards compatible with the existing IBM Quantum Experience hardware support. Second, \textbf{Quintuple} is an ideal quantum computer simulator, but a real quantum computer has a variety of interactions of a quantum register with its environment. Hardware designers attempt to minimize such interactions, but realistically always exist. These interactions, due to the noise of the environment, induce a non-unitary evolution component to the system, resulting in a loss of information called decoherence. The IBM Quantum Experience hardware is no exception to being susceptible to these non-ideal interactions, and it is possible to model these as well in simulation. Doing so will make \textbf{Quintuple} even more useful to researchers designing and implementing algorithms to run on the IBM Quantum Experience, so integrating such modeling is planned for a future update of \textbf{Quintuple}.

\section*{Acknowledgments}
I am currently on leave from the NSF-AAPF grant 1501208 to conduct observations in Antarctica with the South Pole Telescope. I would like to thank Dr. Casey Handmer and Dr. Jerry M. Chow respectively for helpful comments during the preparation of this manuscript. I acknowledge use of the IBM Quantum Experience for this work. The views expressed are those of the author and do not reflect the official policy or position of IBM or the IBM Quantum Experience team. 

\begin{appendices}

\section{Reordering Algorithm}
\label{reordering}
The reordering algorithm allows the user to compare an ordered set of qubits to the output of the quantum computation. Such an algorithm is necessary as while internal to the quantum computer abstraction qubits can be in arbitrary order grouped in arbitrary quantum registers, the user desires output in a specified order. This section presents the details of this algorithm.

There are some requested configurations which would be impossible to provide without merging quantum registers, and the first step of the reordering algorithm computes and merges quantum registers as needed so that it may be possible to sort and return the desired configuration. For example, if the user requests the order ``q0'',``q1'',``q3'',``q4'' and internally ``q0'' and ``q4'' are members of a single quantum register, and ``q1'' and ``q3'' are members of another, these two quantum registers will have to be merged before sorting. 

After the first step, we know we have a set of quantum registers that it is possible to reorder into the requested order. However it could still be the case that we cannot return exactly the requested order. For example if ``q0'',``q1'',``q3'',``q4'' were again requested but in this case internally all 5-qubits reside in the same quantum register, it will be in general not possible to separate out ``q2''. This is checked for and the algorithm throws an exception if sorting is not possible at this stage.

Since we are only dealing with a small number of qubits (5) it is possible to use a simplistic sorting algorithm for clarity; in this case bubble sort is chosen. With each step in the sorting algorithm, we must also rearrange the state of the quantum register involved to correspond to the new order. Using a sorting algorithm with simple well defined operations, bubble sort with its in place swaps, makes it easy to apply the necessary matrix operations to the quantum register.

The bubble sort algorithm is simple to describe: it steps through a list comparing adjacent items and swaps them as necessary, and repeats this stepping through until the list is sorted. It has a worst case performance of $\mathcal{O}(n^2)$. Since $n=5$ in our case this is not a big penalty to pay for simplicity, and the nature of quantum computation makes this the least of our worries were \textbf{Quintuple} attempted to be extended to large $n$. The bubble sort algorithm is explicitly coded so that as we swap the qubits to match the desired order, we also rearrange the state of the quantum register involved to correspond with the new order. This is done by computing the permutation matrix corresponding to the rearranging prescribed by bubble sort, and applying this permutation matrix  to the state. This is done with every swap that bubble sort prescribes of the qubit list, meaning that the state is in the corresponding order when the qubit list is sorted. 

The final step of the reordering algorithm is to just return a single state representing the qubits of interest; this is possible as was ensured in the previous step. For example if ``q0'',``q1'',``q3'',``q4'' are requested, and ``q2'' resides in a separate quantum register than any of these qubits, then ``q2'' is ignored. The result is then easily computed as the ordered tensor product the quantum registers solely containing ordered qubits of interest. This result can be compared to the expected state supplied by the user.

The pseudo-code for the algorithm is included below:
\begin{algorithm}
\caption{Reordering}\label{reorder}
\begin{algorithmic}[1]
\Function{Reorder}{O: requested order}

\Algphase{Phase 1 - Merge quantum registers}
\Require O is in increasing order
\For{$q \in O$} 
  \For{$r \in R \gets \text{ quantum registers}$}
    \State $rmin \gets \text{smallest qubit in r}$
    \State $rmax \gets \text{largest qubit in r}$
    \State $S \gets \text{all qubits between (inclusive) } rmin \text{ and } rmax$
    \If{$q \not \in r \And q \in S $}
      \State $r_q \gets \text{the register q belongs to}$
      \State $\textsc{merge}(r_q,r)$
    \EndIf
  \EndFor
\EndFor
\Algphase{Phase 2 - Sort quantum registers}
\Ensure Every quantum register has qubits that are either all in $O$ or none are in $O$
\For{$r \in R  \gets \text{ quantum registers}$} 
  \State $Q \gets \text{ qubits in }r $ 
  \If{$Q \cap O \not \in \{\emptyset,Q\}$}
    \State \Return $failure$
  \EndIf
  \If{$Q$ not ordered}
    \State $n \gets length(Q)$
    \State $swapped \gets true$
    \While{$swapped \neq  false$}
      \State $swapped \gets false$
      \For{$i=0$ to $n-1$}
        \If{$Q[i]>Q[i+1]$}
            \State $\textsc{Swap}(r,i,i+1)$
            \State $swapped \gets true$
        \EndIf
      \EndFor
    \EndWhile
  \EndIf
\EndFor
\Algphase{Phase 3 - Create combined answer state} 
\State $answer \gets nil$
\For{$r \in R  \gets \text{ quantum registers}$} 
  \State $Q \gets \text{ qubits in }r $ 
  \For{$q \in Q $}
    \If{$Q \in O$}
      \If{$answer = nil$}
        \State $answer \gets q$
      \Else
        \State $answer \gets answer \otimes q$
      \EndIf
    \EndIf
  \EndFor
\EndFor
\State \Return $answer$ 
\EndFunction
\end{algorithmic}
\end{algorithm}

\begin{algorithm}[h]
\caption{Swap }\label{swap}
\begin{algorithmic}[1]
\Procedure{Swap}{$r,i,j$}
\State $Q \gets \text{ qubits in }r $ 
\State $state \gets \text{ state of } r $ 
\Algphase{Phase 1 - Permute the state }
\State $n \gets length(Q)$
\State $L\gets \text{ all possible states of } n \text{ qubits in canonical ordering} $
\State $permute \gets Id_{n\times n}$ \Comment{$n\times n$ identity matrix}
\State $swapped \gets \emptyset $
\For{$c \in L$}
  \State $newc \gets c$
  \State $\textsc{SwapHelper}(newc,i,j)$
  \If{$newc \neq c$}
    \State $i_{\text{per}} \gets \text{ index of } c \text{ in L }$
    \State $j_{\text{per}} \gets \text{ index of } newc \text{ in L }$
    \State $swap \gets \{i_{\text{per}},j_{\text{per}}\}$
    \If{$swap \not \in swapped$}
      \State $swapped \gets swapped \cup swap $
      \State $\textsc{SwapHelper}(permute.rows,i_{\text{per}},j_{\text{per}})$
    \EndIf
  \EndIf
\EndFor
\State $state \gets permute \cdot state$
\Algphase{Phase 2 - Swap the qubits in the register}
\State $\textsc{SwapHelper}(Q,i,j)$
\EndProcedure
\end{algorithmic}
\end{algorithm}

\begin{algorithm}
\caption{Swap Helper}\label{swaphelper}
\begin{algorithmic}[1]
\Procedure{SwapHelper}{$l,i,j$}
\State $tmp \gets l[i]$
\State $l[i] \gets l[j]$
\State $l[j] \gets tmp$
\EndProcedure
\end{algorithmic}
\end{algorithm}
\clearpage

\end{appendices}






\bibliographystyle{phy-bstyles/cpc}
\bibliography{Quintuple}






\end{document}